\documentclass[aps,prl,twocolumn,showpacs,superscriptaddress,floatfix,10pt]{revtex4-1}
\usepackage{framed}
\usepackage{graphicx}
\usepackage{amsmath}
\usepackage{epsfig}
\usepackage{helvet}
\usepackage{amssymb}
\usepackage{framed}
\usepackage{bbold}
\usepackage{subfigure}

\newcommand{\bra}[1]{\mbox{$\langle #1 |$}}
\newcommand{\ket}[1]{\mbox{$| #1 \rangle$}}

\def\LAT{\mbox{\scriptsize lattice}}

\def\cM{\Lambda}
\allowdisplaybreaks[3]

\def\appendix{Appendix }

\def\cMPS{\mbox{\tiny cMPS}}
\def\cMERA{\mbox{\tiny cMERA}}
\def\CFT{\mbox{\tiny CFT}}

\def\UV{\mbox{\tiny UV}}

\begin{document}
\title{Magic entanglement renormalization for quantum fields}

\author{Yijian Zou}
\affiliation{Perimeter Institute for Theoretical Physics, 31 Caroline Street North, Waterloo, ON N2L 2Y5, Canada}
\affiliation{University of Waterloo, Waterloo ON, N2L 3G1, Canada}
\author{Martin Ganahl}
\affiliation{Perimeter Institute for Theoretical Physics, 31 Caroline Street North, Waterloo, ON N2L 2Y5, Canada}
\author{Guifre Vidal}

\affiliation{Perimeter Institute for Theoretical Physics, 31 Caroline Street North, Waterloo, ON N2L 2Y5, Canada}
\affiliation{Alphabet (Google) X, Mountain View, CA 94043, USA}

\begin{abstract}
Continuous tensor networks are variational wavefunctions proposed in recent years to efficiently simulate quantum field theories (QFTs). Prominent examples include the continuous matrix product state (cMPS) 
and the continuous multi-scale entanglement renormalization ansatz (cMERA). 
While the cMPS can approximate ground states of a class of QFT Hamiltonians that are both \textit{local} and \textit{interacting},  cMERA is only well-understood for QFTs that are \textit{quasi-local} and \textit{non-interacting}. 
In this paper we propose the \textit{magic} cMERA, a concrete realization of cMERA for a free boson QFT that
simultaneously satisfies four remarkable properties: 
(i) it is the \textit{exact} ground state of a \textit{strictly local} Hamiltonian;
(ii) in the massless case, its spectrum of scaling operators is exactly soluble in real space;
(iii) it has the short-distance structure of a cMPS; 
(iv) it is generated by a quasi-local entangler that can be written as a continuous matrix product operator. None of these properties is fulfilled by previous cMERA proposals. Properties (iii)-(iv) establish a firm connection between cMERA and cMPS wave-functionals, opening the path to applying powerful cMPS numerical techniques, valid for interacting QFTs, also to cMERA calculations.
\end{abstract}

\maketitle

Tensor networks \cite{Fannes, White, Vidal, Perez-Garcia, MERA, MERA2, MERAalgorithms, criticalMERA, PEPS1, PEPS2, PEPS3, rev1, rev2, rev3, rev4, rev5} offer an efficient representation of otherwise exponentially complex objects, such as the manybody wavefunction of a quantum system. Over the last two decades, tensor networks have become of importance in an ever growing number of research areas, including quantum information and condensed matter \cite{Fannes, White, Vidal, Perez-Garcia, MERA, MERA2, MERAalgorithms, criticalMERA, PEPS1, PEPS2, PEPS3},  quantum chemistry \cite{QC1, QC2, QC3, QC4}, statistical mechanics \cite{CTMRG, TRG, TEFRG, TNR}, quantum gravity \cite{Swingle, dS1, dS2, dS3, MERAgeometry}, and machine learning \cite{ML1, ML2, ML3, ML4, ML5}.  
Two particularly useful tensor networks are the \textit{matrix product state} (MPS) \cite{Fannes, White, Vidal, Perez-Garcia}, successful at representing ground states of gapped Hamiltonians in one spatial dimension, and the \textit{multi-scale entanglement renormalization ansatz} (MERA) \cite{MERA, MERA2, MERAalgorithms, criticalMERA}, specially suited for ground states of critical Hamiltonians. These tensor networks are defined on the lattice. More recently continuous versions that operate directly in the continuum, namely the continuous MPS (cMPS) \cite{cMPS1, cMPS2, cMPS3, cMPS4, cMPS5, cMPS6, cMPS7, cMPS8, cMPS9, cMPS10, cMPS11, cMPS12, cMPS13, cMPS14} and the continuous MERA (cMERA) \cite{cMERA1, Cotler1, Qi, Adrian, Cotler2, Fernandez, cMERA2, cMERA3, cMERA4, cMERA5, cMERA6, cMERA7, cMERA8, cMERA9, cMERA10, Janet}, have also been proposed. They aim to repeat, now for quantum field theories in the continuum, the success of tensor networks for lattice systems. 

Consider for concreteness a bosonic quantum field $\phi(x)$ on the real line $x \in \mathbb{R}$, with conjugate momentum $\pi(x)$ such that $[\phi(x),\pi(y)] = i\delta(x-y)$. Let  
\begin{equation}
\psi(x) \equiv \sqrt{\frac{\Lambda}{2}} \phi(x) + \frac{i}{\sqrt{2\Lambda}}\pi(x)
\end{equation}
denote the annihilation operator and $\ket{\Lambda}$ the vacuum state, with $\psi(x) \ket{\Lambda} =0$ for all $x$. We will focus on the cMERA for this bosonic field, as proposed by Haegeman, Osborne, Verschelde, and Verstraete in Ref. \cite{cMERA1}. It reads
\begin{equation} \label{eq:cMERA}
\ket{\Psi^{\Lambda}(s)} \equiv \mathcal{P}e^{-i\int_0^s ds' \left[L + K(s')\right]} \ket{\Lambda},
\end{equation}
where $\mathcal{P}e$ is a path ordered exponential \textit{in scale}, 
\begin{equation}
L\equiv  \int dx\left[\pi(x) x\partial_x \phi(x) +  \frac{1}{2} \pi(x)\phi(x)\right]
\end{equation}
is the non-relativistic dilation operator, and $K(s)$ is the so-called \textit{entangler}, a \textit{quasi-local} \cite{quasilocal} operator that contains the cMERA variational parameters. Physically, the cMERA describes an entangling evolution in scale $s$, generated by $L+K(s)$, that transforms the unentangled vacuum $\ket{\Lambda}$ into the entangled state $\ket{\Psi^{\Lambda}(s)}$. A simple example of (quadratic, scale-independent) entangler is \cite{cMERA1}
\begin{equation} \label{eq:K}
K \equiv \frac{-i}{2}\int dxdy~ g(x-y) \left[\psi(x)\psi(y) - \psi(x)^{\dagger}\psi(y)^{\dagger}\right].
\end{equation}
Here $g(x)$ is some smearing function (\textit{e.g.} $g(x)\propto e^{-\sigma(\Lambda x)^2/4}$ \cite{sigma}) with a built-in UV length scale $1/\Lambda$ such that, importantly, no entanglement is introduced in $\ket{\Psi^{\Lambda}(s)}$ at distances shorter than $1/\Lambda$. Example \eqref{eq:K} is independent of scale, $K(s) = K$. Then \eqref{eq:cMERA} simplifies to
\begin{equation}\label{eq:cMERAK}
\ket{\Psi^{\cM}(s)} =  e^{-is(L+K)} \ket{\Lambda},
\end{equation}
and in the limit of a large entangling evolution we obtain
\begin{equation} \label{eq:cMERAQi}
\ket{\Psi^{\Lambda}} \equiv \lim_{s\rightarrow \infty} e^{-is(L+K)} \ket{\Lambda},
\end{equation}
namely a fixed-point, scale-invariant cMERA $\ket{\Psi^{\Lambda}}$ \cite{Qi}. The generator $L+K$ is a quasi-local version of the dilation operator in a \textit{conformal field theory} (CFT) and comes with its own spectrum of smeared scaling operators $\mathcal{O}_{\alpha}^{\Lambda}(x)$ and related conformal data \cite{Qi}.

To date, the cMERA is only well-understood when, as in the above example \eqref{eq:K}, the entangler is \textit{quadratic} in the fields, thus representing a \textit{Gaussian} wavefunctional that describes ground states of \textit{non-interacting} Hamiltonians (see also \cite{Gaussian} for first steps beyond Gaussian cMERA). Clearly, unleashing the true potential of cMERA will require the discovery of non-perturbative algorithms for \textit{interacting} QFTs. We envisage that those algorithms will eventually match the power of existing lattice MERA algorithms \cite{MERAalgorithms}. In the meantime, however, the Gaussian cMERA for non-interacting QFTs is already of significant value in its own right. It offers an explicit demonstration that MERA can be generalized from the lattice to the continuum, and a framework for both building toy models of holography in quantum gravity \cite{cMERA2, cMERA3, cMERA4, cMERA5, cMERA6, cMERA7, cMERA8, cMERA9, cMERA10} and studying QFTs in curved spacetime \cite{Janet}.

\begin{figure}
\includegraphics[width=0.49\linewidth]{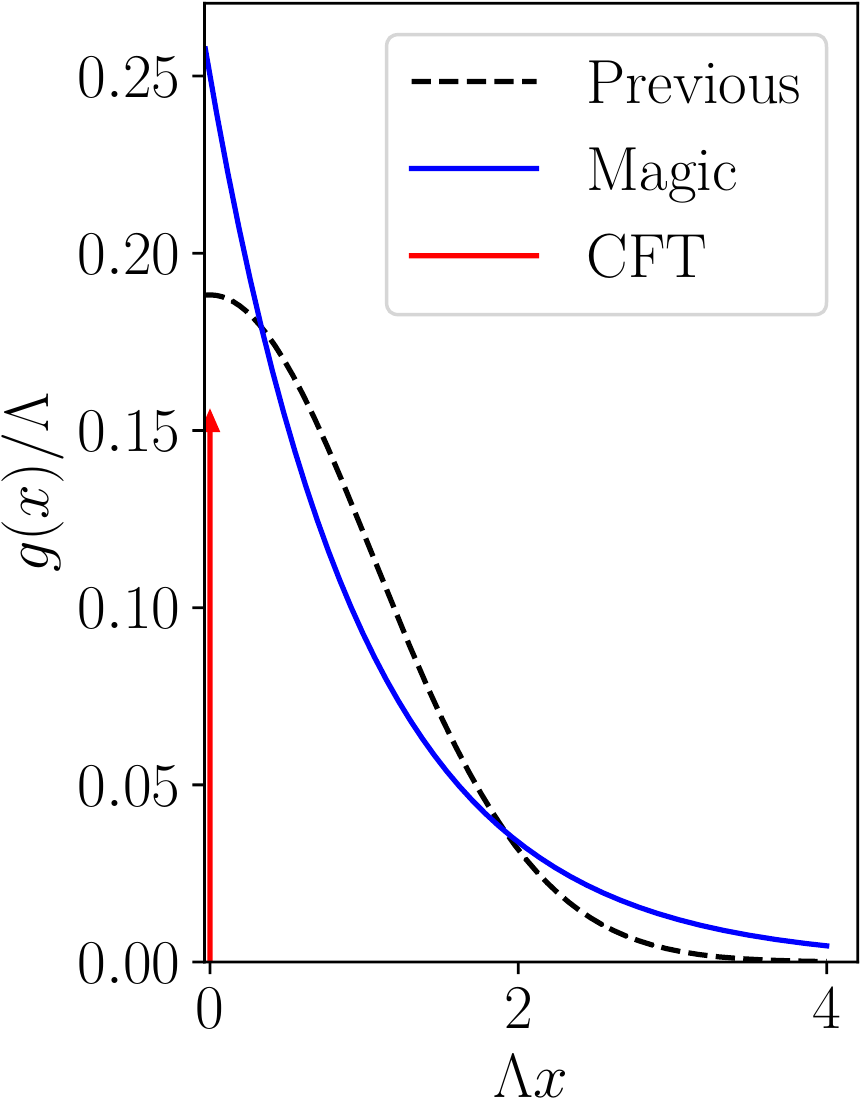}
\includegraphics[width=0.49\linewidth]{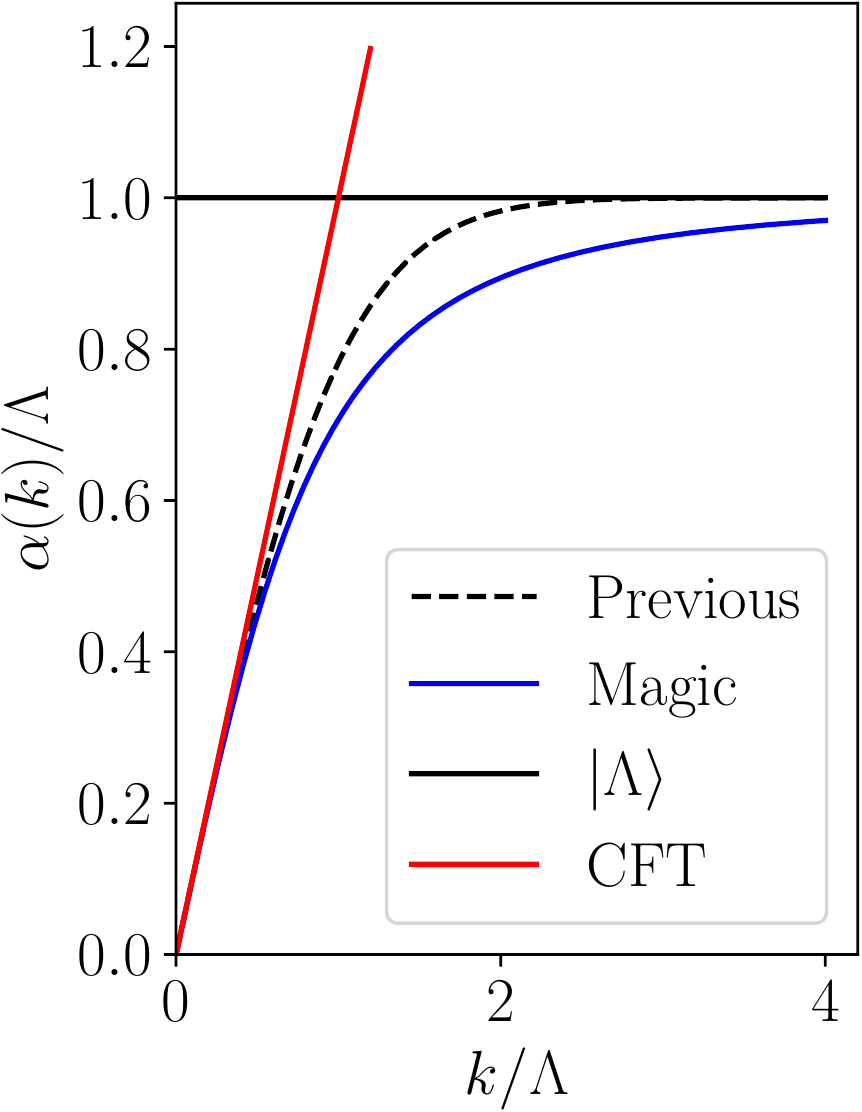}
\caption{
(Left) Smearing function $g(x)$ of the quadratic entangler $K$ in \eqref{eq:K}. The exponential $\Lambda e^{-\Lambda|x|}/4$ of the magic cMERA produces the ground state of a local Hamiltonian such as \eqref{eq:HLamCFT} and \eqref{eq:HLam2}. In contrast, the Gaussian $(\Lambda/4) \sqrt{\sigma/\pi}~e^{-\sigma(\Lambda x)^2/4}$ of previous proposals \cite{cMERA1,Qi,sigma} results in the ground state of a quasi-local Hamiltonian \cite{Qi}. The CFT ground state corresponds to a delta distribution $\delta(x)$, which is the limit $\Lambda \rightarrow \infty$ of both smearing functions.
(Right) Function $\alpha(k)$ in \eqref{eq:alpha} that defines the magic cMERA annihilation operators $a^{\Lambda}(k)$ in \eqref{eq:aLam}. At small $k\ll\Lambda$ (large distances) it approaches $|k|$, corresponding to a CFT, whereas at large $k \gg \Lambda$ (short distances) it approaches the constant $\Lambda$, corresponding to the unentangled vacuum $\ket{\Lambda}$. 
\label{fig:magic}
}
\end{figure}

In this paper we propose a concrete realization of cMERA, dubbed \textit{magic} cMERA, specified by the following choice of smearing function in \eqref{eq:K}:
\begin{equation} \label{eq:gx}
g(x) \equiv \frac{\Lambda}{4} e^{-\Lambda |x|}~~~~(\mbox{magic cMERA})
\end{equation}
see Fig. \ref{fig:magic}(a). We will show that the magic cMERA simultaneously fulfils four remarkable properties. (i) The state $\ket{\Psi^{\Lambda}(s)}$ in \eqref{eq:cMERAK} is the exact ground state of a \textit{strictly local} Hamiltonian (previous cMERA realizations were the ground state of a Hamiltonian that was, at best, quasi-local \cite{cMERA1, Qi}). 
(ii) The scale-invariant $\ket{\Psi^{\Lambda}}$ in \eqref{eq:cMERAQi} has scaling operators $\mathcal{O}_{\alpha}^{\Lambda}(x)$ whose real-space profile can be solved \textit{exactly}  (previous examples required a numerical Fourier transform \cite{Qi}). 
(iii) At short distances, the magic cMERA has the same entanglement structure as a cMPS (while previous proposals are seen to be UV-inequivalent to a cMPS). 
(iv) The corresponding entangler $K$ is efficiently represented as a simple \textit{continuous matrix product operator} (cMPO). 
Results (i)-(ii) connect the cMERA formalism to local Hamiltonians while providing important analytical insight into the real-space structure of entanglement renormalization for quantum fields. In turn, results (iii)-(iv) establish an intriguing, direct connection between cMERA and cMPS formalisms, paving the way to using cMPS techniques, valid for interacting QFTs, in cMERA calculations. Our work thus sets the foundations for a much anticipated, real-space computational framework for cMERA, 
analogous to existing lattice MERA algorithms \cite{MERAalgorithms}, that we 
further develop in Ref. \cite{Martin}.

\textit{Exact ground state of a local Hamiltonian}.---We start by considering the fixed-point cMERA $\ket{\Psi^{\Lambda}}$, given by \eqref{eq:cMERAQi} with the magic smearing profile $g(x)$ in \eqref{eq:gx}. We claim that $\ket{\Psi^{\Lambda}}$ is the exact ground state of the Hamiltonian
\begin{equation} \label{eq:HLamCFT}
H^{\cM} \equiv \frac{1}{2}\!\int\! dx \left( \!\pi(x)^2 + (\partial_x \phi(x))^2  + \frac{1}{\Lambda^2} \left(\partial_x \pi(x)\right)^2\right)\!.
\end{equation}
This \textit{local} Hamiltonian can be readily interpreted as the free boson CFT Hamiltonian 
\begin{equation} \label{eq:HCFT}
H^{\CFT} \equiv \frac{1}{2}\int dx\left( \pi(x)^2 + (\partial_x \phi(x))^2 \right)
\end{equation}
describing a \textit{relativistic} massless boson, modified at small distances by a \textit{non-relativistic} UV regulator 
\begin{equation}
A^{\cM}_{\UV} \equiv \frac{1}{2\Lambda^2} \int dx(\partial_x \pi(x))^2,
\end{equation}
that is $H^{\cM} =  H^{\CFT} + A^{\cM}_{\UV}$. 
To prove the above claim, we first introduce a complete set of annihilation operators,
\begin{equation} \label{eq:aLam}
a^{\Lambda}(k) \equiv \sqrt{\frac{\alpha(k)}{2}} \phi(k) + \frac{i}{\sqrt{2\alpha(k)}} \pi(k),
\end{equation}
in terms of the Fourier space modes
\begin{equation}
\phi(k) \equiv \int\! dx \frac{e^{-ikx}}{\sqrt{2\pi}} \phi(x),~~\pi(k) \equiv \int\! dx \frac{e^{-ikx}}{\sqrt{2\pi}} \pi(x),
\end{equation}
and of the function (see Fig. \ref{fig:magic}(b))
\begin{equation} \label{eq:alpha}
\alpha(k) \equiv \sqrt{\frac{k^2 \Lambda^2}{k^2+\Lambda^2}}. 
\end{equation}
Function $\alpha(k)$ relates to the Fourier transform \cite{Qi}
\begin{equation}
g(k) \equiv \int dx~e^{ikx} g(x) = \frac{\Lambda^2}{2 (k^2 + \Lambda^2)}
\end{equation}  
of the smearing function $g(x)$ through $d \alpha(k)/d k = 2\alpha(k)g(k)/k$.
Ref. \cite{Qi} showed that an alternative characterization of the fixed-point $\ket{\Psi^{\Lambda}}$ is then as the state that is simultaneously annihilated by all the $a^{\Lambda}(k)$'s  
\begin{equation} \label{eq:aPsi}
a^{\Lambda}(k) \ket{\Psi^{\Lambda}} = 0 ,~~~\forall k\in \mathbb{R}.
\end{equation}
Now, writing $H^{\Lambda}$ also in terms of $a^{\Lambda}(k)$, we obtain \cite{Supp}
\begin{equation} \label{eq:diagHLamCFT}
 H^{\Lambda} = \int\! dk ~E^{\Lambda}(k) ~a^{\Lambda}(k)^{\dagger}~a^{\Lambda}(k),  
\end{equation} 
with single-particle energies $E^{\Lambda}(k) = |k| \sqrt{1+\left(\frac{k}{\Lambda}\right)^2}$. Therefore $\ket{\Psi^{\Lambda}}$ is indeed the ground state of $H^{\Lambda}$. Notice that the UV regulator $A^{\Lambda}_{\UV}$ introduces $O\left(|k|^3/\Lambda^2\right)$ corrections to the CFT dispersion relation $E^{\CFT}(k) = |k|$, which are negligible at low energies.
 
 \begin{figure}
\includegraphics[width=0.49\linewidth]{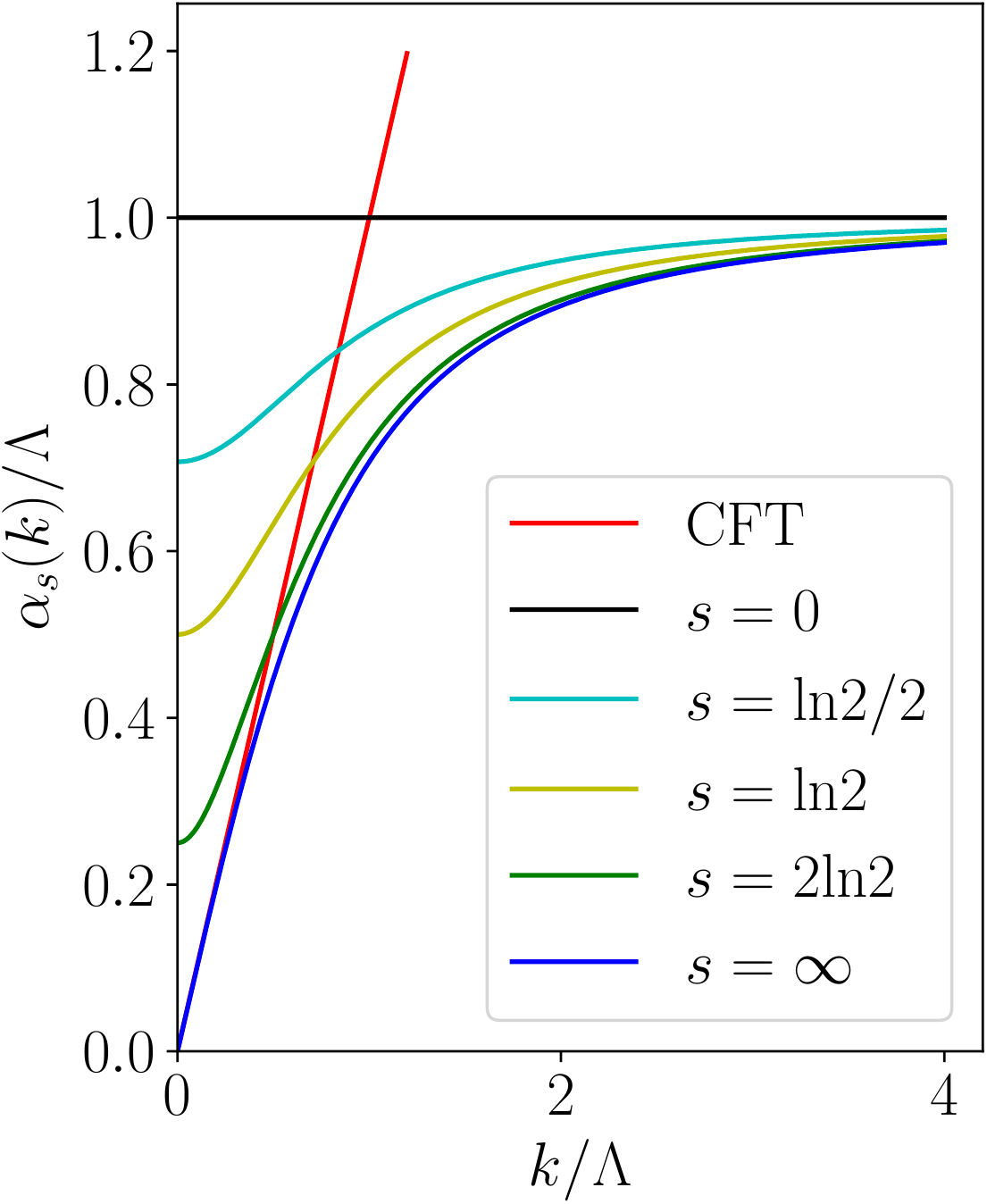}\includegraphics[width=0.49\linewidth]{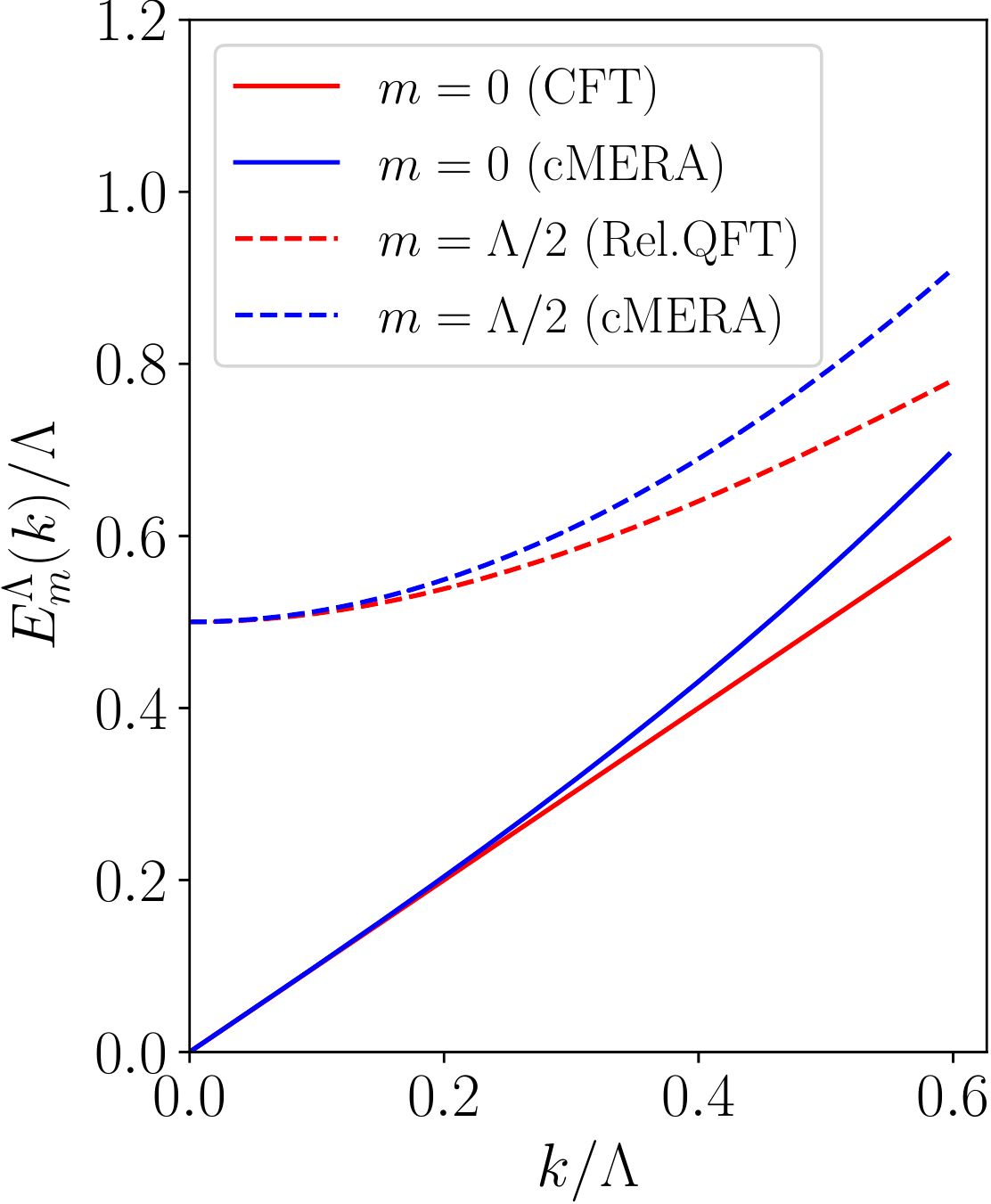}
\caption{
(Left) Function $\alpha_s(k)$ in \eqref{eq:alphas} for a sequence of increasingly large values of the scale parameter $s$. We see that as $s$ increases, $\alpha_s(k)$ smoothly interpolates between the unentangled vacuum $\ket{\Lambda}$ at $s=0$ and the fixed-point cMERA $\ket{\Psi^{\Lambda}}$ in the limit $s \rightarrow \infty$.
(Right) Single-particle energies $E^{\Lambda}_m(k)$ for the massless case $m=0$ (or $s=\infty$) and the massive case $m= \Lambda/2$ (or $s=\log 2$). For comparison, we also show (in red) the dispersion relation corresponding to the CFT and relativistic massive QFT.
\label{fig:sfinite}
}
\end{figure}

We emphasize how surprising this first result is. It is indeed highly non-trivial that a choice of smearing function $g(x)$ actually exists for which the exact fixed-point \eqref{eq:cMERAQi} of a \textit{quasi-local} generator (namely $L+K$) is at the same time the ground state of a \textit{strictly local} Hamiltonian. Moreover, a similar derivation \cite{Supp} shows that the magic cMERA $\ket{\Psi^{\Lambda}(s)}$ in \eqref{eq:cMERAK} resulting from a \textit{finite} entangling evolution in scale and annihilated by 
\begin{eqnarray}\label{eq:aLams}
a^{\cM}_s(k) &\equiv& \sqrt{\frac{\alpha_s(k)}{2}} \phi(k) + \frac{i}{\sqrt{2\alpha_s(k)}}\pi(k), ~~~~~~~\\
\alpha_s(k) &\equiv& \sqrt{\frac{\Lambda^2k^2 + \Lambda^4 e^{-2s} }{k^2 + \Lambda^2}} \label{eq:alphas}
,~~~~
\end{eqnarray} 
(see Fig. \ref{fig:sfinite}(a) for the evolution of $\alpha_{s}(k)$ with $s$), 
is the ground state of the gapped \textit{local} Hamiltonian $H^{\Lambda}_m$,
\begin{equation}
H^{\Lambda}_m \equiv H^{\Lambda} + \frac{1}{2}\int dx~m^{2}\phi(x)^2,~~~~~~m\equiv \Lambda e^{-s}. \label{eq:HLam2}
\end{equation}
Its dispersion relation $E_m(k) = \sqrt{|k|^2+m^2} \sqrt{1+\left(\frac{k}{\Lambda}\right)^2}$ is shown in Fig. \ref{fig:sfinite}(b). 
This Hamiltonian can be written as $H_m^{\Lambda} = H^{\CFT} + \frac{1}{2}\int dx~m^{2}\phi(x)^2 + A^{\Lambda}_{\UV}$, and thus be interpreted for $m \ll \Lambda$ as that of a \textit{relativistic} boson with finite mass $m$ (IR regulator), modified again at short distances by the non-relativistic term $A_{UV}^{\Lambda}$ (UV regulator). An \textit{exhaustive} characterization of cMERA ground states for \textit{local quadratic} Hamiltonians can be found in \cite{Supp}.

\begin{figure}
\includegraphics[width=0.49\linewidth]{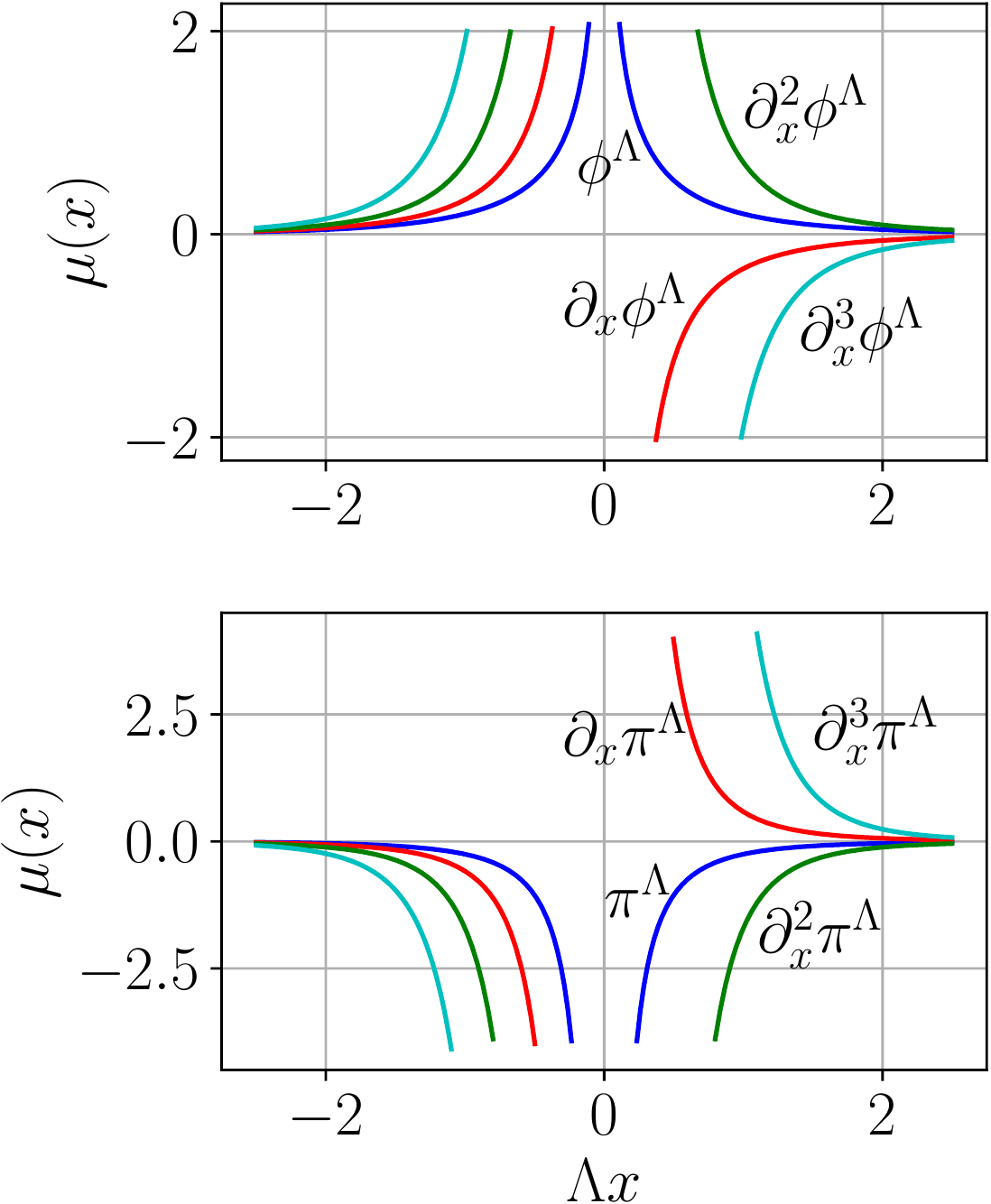}
\includegraphics[width=0.49\linewidth]{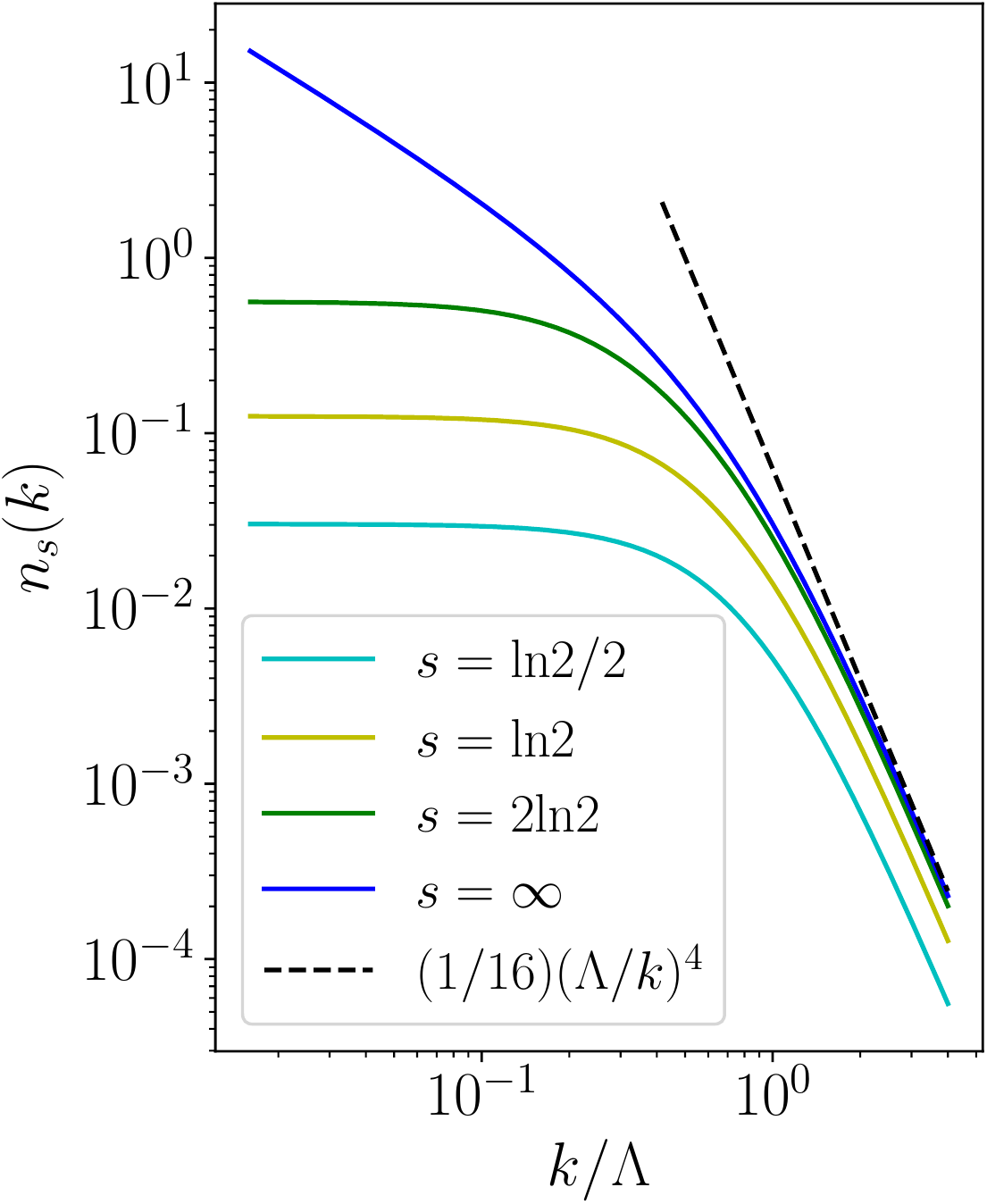}
\caption{
(Left) Profile functions for $\phi^{\Lambda}$ and $\pi^{\Lambda}$, see first term in \eqref{eq:phireal} and \eqref{eq:pireal}, as well as some of its derivative descendants. 
(Right)  Correlator $n_s(k)$ in \eqref{eq:nk} for the magic cMERA $\ket{\Psi^{\Lambda}(s)}$ for different values of $s \in [0,\infty)$. For any non-zero amount of entangling evolution, $s>0$, the magic cMERA state displays UV correlations $\sim (\Lambda/k)^4$. A cMPS optimized for the ground state of $H^{\Lambda}_m$ for $m=\Lambda e^{-s}$ (not shown) was also computed and seen to accurately reproduce the same $n_s(k)$. 
\label{fig:nk} 
} 
\end{figure}

\textit{Exactly soluble scaling operators.---} Ref. \cite{Qi} showed that the massless boson cMERA can be interpreted as a quasi-local realization of the conformal group. Specifically, the generator $L+K$ corresponds to a smeared version of the dilation operator $D^{\CFT}$ of the free boson CFT. One can then characterize the corresponding smeared scaling operators $\mathcal{O}^{\Lambda}_{\alpha}(x)$ (at $x=0$) as the solutions of 
\begin{equation}
 -i\left[ L+K, \mathcal{O}_{\alpha}^{\Lambda} (0)\right] = \Delta_{\alpha} \mathcal{O}^{\Lambda}_{\alpha}(0).
\end{equation}
This equation has been solved \cite{Qi}. For instance, the scaling operators $\phi^{\Lambda}(x)$ and $\pi^{\Lambda}(x)$ with scaling dimension $\Delta_{\phi}=0$ and $\Delta_{\pi}=1$ \cite{phiscaling}, are the Fourier transforms of
\begin{equation}
\phi^{\Lambda}(k) \equiv \sqrt{\frac{\alpha(k)}{|k|}}\phi(k),~~~\pi^{\Lambda}(k) \equiv \sqrt{\frac{|k|}{\alpha(k)}} \pi(k).
\end{equation}
However, for a generic smearing function $g(x)$ (e.g. $g(x)\propto e^{-\sigma(\Lambda x)^2}$ in Refs. \cite{cMERA1,Qi}) the real-space expression for $\phi^{\Lambda}(x)$ and $\pi^{\Lambda}(x)$ is the sum of two terms \cite{Qi},
\begin{eqnarray}
\phi^{\Lambda}(x) &=& \frac{2^{\frac{3}{4}}\Lambda}{\Gamma(\frac{1}{4})}\int dy ~\frac{K_{\frac{1}{4}}\left( \Lambda|x-y|\right)}{|\Lambda (x-y)|^{\frac{1}{4}}} \phi(y) + \cdots, ~~~ \label{eq:phireal}\\
\pi^{\Lambda}(x) &=& \frac{2^{\frac{5}{4}}\Lambda}{\Gamma(-\frac{1}{4})}\int dy ~\frac{K_{\frac{3}{4}}\left(\Lambda|x-y|\right)}{|\Lambda (x-y)|^{\frac{3}{4}}} \pi(y) + \cdots, ~~~\label{eq:pireal}
\end{eqnarray}
where $K_n$ is the modified Bessel function of the second kind, $\Gamma$ is the Euler gamma function, and $+ \cdots$ denotes the omitted second terms, for which no analytical expression is known. Again rather surprisingly, with the magic smearing function $g(x)$, the second term in both \eqref{eq:phireal} and \eqref{eq:pireal} conveniently vanish and we are just left with the analytical part of the solution, which is plotted in Fig. \ref{fig:nk}(a) for $\phi^{\Lambda}(x)$, $\pi^{\Lambda}(x)$, and some of their derivative descendants. These exact expressions are important when developing a real-space cMERA algorithm \cite{Martin}, since they allows us to compare numerical and exact solutions.

\textit{Same UV entanglement structure as the cMPS.---} A third intriguing, unexpected aspect of the magic cMERA is that it matches the structure of correlations of a cMPS at short distances. The evidence for this is two-fold. On the one hand, we can rewrite $H^{\Lambda}_m$ in \eqref{eq:HLam2} as
\begin{eqnarray}
 H^{\Lambda}_m &=& \int dx  \left( \frac{1}{\Lambda} \partial_x \psi^{\dagger}(x)\partial_x \psi(x) + \frac{\Lambda^2 + m^2}{2\Lambda}\psi^{\dagger}(x)\psi(x)\right. \nonumber \\
&-&        \left. \frac{\Lambda^2 - m^2}{4\Lambda} \left(\psi(x)^2 + \psi(x)^{\dagger 2}\right) \right).~~~\label{eq:HLam}
\end{eqnarray}
The first contribution is the non-relativistic kinetic term
\begin{equation}
 T \equiv \int dx  ~ \frac{1}{\Lambda} \partial_x \psi^{\dagger}(x)\partial_x \psi(x)
\end{equation}
present also in all Hamiltonians for which the cMPS has been seen to accurately describe the ground state, such as the Lieb-Liniger model \cite{cMPS1, cMPS2, cMPS3, cMPS4, cMPS5, cMPS6, cMPS7, cMPS8, cMPS9, cMPS10, cMPS11, cMPS12, cMPS13, cMPS14}. In other words, both cMPS and magic cMERA appear to describe the ground state of Hamiltonians dominated at high energies/short distances by $T$. On the other hand, the UV entanglement structure of both cMPS and magic cMERA can be directly probed by the momentum space correlator 
\begin{equation} \label{eq:nk}
n(k,q) \equiv \langle \psi^{\dagger}(k)\psi(q)\rangle = \delta(k-q) n(k).
\end{equation}
It is known that a generic cMPS satisfies $n_{\cMPS}(k) \sim 1/k^4$ for large $k$ \cite{cMPS2}. Considering for simplicity the fixed-point cMERA \eqref{eq:cMERAQi}, Eq. \eqref{eq:aPsi} can be used \cite{Supp} to also find   
\begin{equation}
n_{\cMERA}(k ) = \frac{\Lambda}{4\alpha(k)} + \frac{\alpha(k)}{4\Lambda} - \frac{1}{2} \sim \frac{1}{k^4},~~~k \gg \Lambda,
\end{equation}
see Fig. \ref{fig:nk}(b). The similar scaling at large $k$ strongly suggests again that the short distance properties of magic cMERA may be well approximated by a cMPS. This is numerically confirmed and then exploited in Ref. \cite{Martin}. 

Notice that on a lattice system all states have the same UV structure: they are all vectors in the tensor product $\bigotimes_j \mathcal{H}_j$ of local vector spaces $\mathcal{H}_j$ describing the lattice sites $j$. Thus, MERA and MPS are trivially UV-equivalent on the lattice. In the continuum, however, two wavefunctionals can differ very significantly at short distances. For instance, the cMERA realization of Refs. \cite{cMERA1, Qi} with the smearing function $g(x) \propto e^{-\sigma(\Lambda x)^2/4}$ leads to a $n(k)$ that decays with large $k$ faster than any power \cite{Supp} and it is therefore UV-inequivalent to a cMPS. It is thus a third non-trivial result that the magic cMERA matches the UV structure of the cMPS.

\textit{Continuous matrix product entangler $K$.---} Finally, the quasi-local entangler $K$ in \eqref{eq:cMERAK} with our choice $g(x)$ in \eqref{eq:gx},
\begin{equation} \label{eq:Kg}
K = \frac{-i\Lambda}{8} \! \int \!\!dx dy~ e^{-\Lambda|x-y|} \left[\psi(x)\psi(y) - \psi(x)^\dagger \psi(y)^\dagger\right],~~~
\end{equation}
can be written as a simple continuous matrix product operator (cMPO) \cite{Supp}. This is best seen by first discretizing the real line, with a regular lattice whose sites are at positions $x_m \equiv \epsilon m$ for $m\in \mathbb{Z}$, with $\epsilon \ll 1/\Lambda$, where $\epsilon$ is the lattice spacing, and introducing a set of lattice annihilation operators  $b_m \approx \sqrt{\epsilon} \psi(x_m)$, with $\left[b_m, b_n^{\dagger} \right] = \delta_{mn}$. Then a lattice approximation $K^{\LAT}$ of entangler $K$ is
\begin{equation}
K^{\LAT} \equiv  \frac{-i\Lambda \epsilon}{4} \sum_{m>n} e^{-\Lambda \epsilon |m-n|} \left[ b_mb_n - b_{m}^{\dagger}b_{n}^{\dagger}\right],
\end{equation}
which can be written as a MPO with matrices $A_m$ \cite{Supp}, 
\begin{equation}
A_m \equiv 
  \left(
    \begin{array}{cccc}
      \mathbb{1} &  \sqrt{\beta e^{-\Lambda\epsilon}\epsilon } ~  b_m^{\dagger} &  \sqrt{\beta e^{-\Lambda\epsilon}\epsilon } ~ b_m & 0\\
      0 & e^{-\Lambda\epsilon}\mathbb{1} & 0 & \sqrt{\beta e^{-\Lambda\epsilon}\epsilon } ~ b_m^{\dagger}\\
      0 & 0 & e^{-\Lambda\epsilon}\mathbb{1} & \sqrt{\beta e^{-\Lambda\epsilon}\epsilon } ~ b_m\\
      0&0&0&\mathbb{1} 
    \end{array} \right),
    \label{eq:MPO}
\end{equation}
namely $K^{\LAT} = \bra{1}A_1 A_2 \cdots A_N\ket{4}$ \cite{Supp}.
To return to the continuum, we reintroduce the field operators $\psi(x_m) = b_m/\sqrt{\epsilon}$ and expand $A_m = \mathbb{1} + \epsilon \mathcal{A}(x_m) + O(\epsilon^2)$. The cMPO matrix $\mathcal{A}(x)$ then reads \cite{Martin,Supp}
\begin{equation}
\mathcal{A}(x) \equiv 
  \left(
    \begin{array}{cccc}
      0 & \sqrt{\beta} \psi^{\dagger}(x) & \sqrt{\beta}  \psi(x) & 0\\
      0 & -\Lambda\mathbb{1} & 0 & \sqrt{\beta}  \psi^{\dagger}(x)\\
      0 & 0 & -\Lambda\mathbb{1} & \sqrt{\beta}  \psi(x)\\
      0&0&0&0 
    \end{array} \right), 
    \label{eq:cMPO}
\end{equation}
and $K = \bra{1} \mathcal{P}e^{ \int dx ~\mathcal{A}(x)} \ket{4}$, where the path ordered exponential is defined by
\begin{equation}
\mathcal{P}e^{ \int dx ~\mathcal{A}(x)} \equiv \! \lim_{\scriptsize{\begin{array}{c} \epsilon\rightarrow 0\\N \rightarrow \infty \end{array}}}\left(\mathbb{1} + \epsilon \mathcal{A}_1 \right) \left(\mathbb{1} + \epsilon \mathcal{A}_2 \right) \cdots \left(\mathbb{1} + \epsilon \mathcal{A}_N \right).
\end{equation}
Here $L\equiv N \epsilon$ is kept finite in taking the double limit $\epsilon \rightarrow 0$ and $N \rightarrow \infty$.
The above cMPO allows us to efficiently apply $K$, as well as its exponential $e^{is K}$ for small $s \ll 1$, to a cMPS and in this way numerically simulate, using well-established cMPS techniques, the scale evolution generated by $L+K$ \cite{Martin}. 

\textit{Discussion.---} We have proposed the \textit{magic} cMERA, a concrete realization of the Gaussian cMERA that, in contrast with previous proposals \cite{cMERA1,Qi}, is the exact ground state of a local Hamiltonian. In addition, it satisfies three other surprising, seemingly unrelated properties: its scaling operators $\mathcal{O}^{\Lambda}_{\alpha}(x)$ can be exactly solved, its short distance entanglement structure equates that of a cMPS, and its quasi-local entangler is compactly encoded by a simple cMPO. 
In Ref. \cite{Martin} we put these properties to work. The entangling evolution in scale of the magic cMERA is simulated with cMPS techniques, even for an interacting entangler. In this way one can efficiently compute expectation values for a class of non-Gaussian wavefunctionals. These wavefunctionals can display power-law decay of correlations and logarithmic scaling on entanglement entropy and are therefore not equivalent to a cMPS. 

The authors thank Qi Hu for many insightful discussions, and acknowledge support from Compute Canada. G. Vidal is a CIFAR fellow in the Quantum Information Science Program. Y. Zou and M. Ganahl thank X for hospitality. X is formerly known as Google[x] and is part of the Alphabet family of companies, which includes Google, Verily, Waymo, and others (www.x.company). Research at Perimeter Institute is supported by the Government of Canada through the Department of Innovation, Science and Economic Development Canada and by the Province of Ontario through the Ministry of Research, Innovation and Science.

\section{Appendices}

\subsection{Appendix I: Exact ground state of Hamiltonian $H^{\Lambda}_m$}

Let us consider a single bosonic quantum field $\phi(x)$ in one spatial dimension, together with its conjugate momentum $\pi(x)$. They obey  the commutation relation $\left[\phi(x),\pi(y) \right] = i\delta(x-y)$. Let us also introduce the annihilation operator $\psi(x)$ and $\psi(k)$ in real and momentum space
\begin{eqnarray}
\psi(x) &\equiv& \sqrt{\frac{\Lambda}{2}}\phi(x)+\frac{i}{\sqrt{2\Lambda}} \pi(x),~~~~\\
\psi(k) &\equiv& \frac{1}{\sqrt{2\pi}}\int\! dx~ e^{-ikx} \psi(x)
\end{eqnarray}
with $\left[\psi(x),\psi(y)^{\dagger} \right]= \delta(x-y)$ and $\left[\psi(k),\psi(q)^{\dagger} \right]= \delta(k-q)$.
 In this work we studied the Hamiltonian $H_m^{\Lambda}$,
\begin{widetext}
\begin{eqnarray}
H^{\Lambda}_m &\equiv& \frac{1}{2}\int\!\! dx\left(\frac{1}{\Lambda^2}(\partial_x \pi(x))^2 +\pi(x)^2 + (\partial_x \phi(x))^2 + m^2 \phi(x)^2\right) \label{eq:app:H1}\\
&=&\frac{1}{\Lambda}\int\! dx   \left(  \partial_x \psi^{\dagger}(x)\partial_x \psi(x) + \frac{m^2 + \Lambda^2}{2} \psi^{\dagger}(x)\psi(x) + \frac{m^2 - \Lambda^2}{4} \left(\psi(x)^2 + \psi(x)^{\dagger 2}\right) \right) \label{eq:app:H2} \\
&=&\frac{1}{\Lambda} \int\!\! dk\left(k^2 + \frac{m^2 + \Lambda^2}{2} \right)\psi^{\dagger}(k)\psi(k) + \frac{m^2-\Lambda^2}{4}\left(\psi(k)\psi(-k) + \psi(k)^{\dagger}\psi(-k)^{\dagger}\right) \label{eq:app:H3}
\end{eqnarray}
\end{widetext}
This Hamiltonian can be diagonalized by introducing the annihilation operators
\begin{equation}\label{eq:app:aLams}
a^{\cM}_s(k) \equiv \sqrt{\frac{\alpha_s(k)}{2}} \phi(k) + \frac{i}{\sqrt{2\alpha_s(k)}}\pi(k),
\end{equation}
where the function $\alpha_s(k)$ is given by
\begin{equation}
\alpha_s(k) \equiv \sqrt{\frac{\Lambda^2k^2 + \Lambda^4 e^{-2s} }{k^2 + \Lambda^2}} = \sqrt{\frac{\Lambda^2k^2 + \Lambda^2 m^2 }{k^2 + \Lambda^2}} \label{eq:app:alphas} 
\end{equation}
where $s = \log(\Lambda/m)$. Indeed, the Hamiltonian can be rewritten as
\begin{eqnarray}
H^{\Lambda}_m &=& \int\!\! dk~ E^{\Lambda}_{m}(k) ~ a^{\Lambda}_{s}(k)^{\dagger} a^{\Lambda}_s(k),~~\\\label{eq:app:H4}
E^{\Lambda}_m(k) &\equiv& \sqrt{|k|^2+m^2} \sqrt{1+\left(\frac{k}{\Lambda}\right)^2}\label{eq:app:ELam}
\end{eqnarray}
as can be checked by direct replacement.
Notice that these expressions are valid for any value of the mass parameter $m \in \mathbb{R}$, which we can take to be positive since only $m^2$ appears in the Hamiltonian. Let us first consider two limit cases $m=\Lambda$ and $m=0$ and then the full scale evolution.

\subsubsection{1. Case $m=\Lambda$: unentangled ground states }

For $m=\Lambda$ (or $s=0$) the Hamiltonian simplifies to
\begin{eqnarray}
H_{m=\Lambda}^{\Lambda} &=& \frac{1}{\Lambda} \int\!\! dk (k^2 + \Lambda^2 )\psi^{\dagger}(k)\psi(k) 
\end{eqnarray}

and
\begin{equation}\label{eq:appaLams}
a^{\cM}_{s=0}(k) \equiv \sqrt{\frac{\Lambda}{2}} \phi(k) + \frac{i}{\sqrt{2\Lambda}}\pi(k),
\end{equation}
where the function $\alpha_{s=0}(k)$ is given by
\begin{equation}
\alpha_{s=0}(k) \equiv \sqrt{\frac{\Lambda^2k^2 + \Lambda^4  }{k^2 + \Lambda^2}} = \Lambda. \label{eq:app:alphas2}
\end{equation}
The ground state is the product state $|\Lambda\rangle$ that is annihilated by any $\psi(k)$, that is $\psi(k)\ket{\Lambda}=0~\forall k$, because the Hamiltonian is positive definite. Through a Fourier transform, this condition is equivalent to $\psi(x)\ket{\Lambda}=0 ~~\forall x$, which we used in the main text to define the unentangled vacuum state $\ket{\Lambda}$.

\subsubsection{2. Case $m=0$: critical ground state }
For $m=0$ (or $s=\infty$) the Hamiltonian is gapless. This can be seen by the fact that 
\begin{equation}
\label{eq:APPalphak}
\alpha(k)\equiv \alpha_{s=\infty}(k)=\sqrt{\frac{k^2\Lambda^2}{k^2+\Lambda^2}}
\end{equation}
at small $k\ll \Lambda$ reduces to the CFT profile
\begin{equation}
\alpha(k)=\alpha^{\CFT}(k)-\frac{|k|^3}{2\Lambda^2}+O(k^5),
\end{equation}
where
\begin{equation}
\alpha^{\CFT}(k)=|k|.
\end{equation}
At large $k\gg \Lambda$, the state approaches the unentangled state $\ket{\Lambda}$, in the sense that
\begin{equation}
\alpha(k)=\Lambda-\frac{\Lambda^3}{k^2}+O(k^{-4}).
\end{equation}

The dispersion relation also reduces to the CFT dispersion $E^{\CFT}(k)=|k|$ at small $k\ll \Lambda$. At large $k\gg \Lambda$, the dispersion relation is dominated by the nonrelativistic kinetic energy,
\begin{equation}
E^{\Lambda}(k)=\frac{k^2}{\Lambda}+O(|k|).
\end{equation}
\subsubsection{3. Case $0 < m< \Lambda$: scale evolution}
We will show that the ground state of Eq.~\eqref{eq:app:H1} with $m=\Lambda e^{-s}$ is the cMERA state 
\begin{equation}
\label{eq:APPevol}
\ket{\Psi^{\cM}(s)} =  e^{-is(L+K)} \ket{\Lambda},
\end{equation}
where $K$ is the magic entangler
\begin{equation}
\label{eq:APPK}
K=\frac{-i}{2}\int dk \, g(k)(\psi(k)\psi(-k)-\psi^{\dagger}(k)\psi^{\dagger}(-k))
\end{equation}
with
\begin{equation}
g(k)=\frac{\Lambda^2}{2(k^2+\Lambda^2)}.
\end{equation}
Clearly, at $s=0$ the cMERA state is $|\Lambda\rangle$, the ground state of Eq.~\eqref{eq:app:H1} with $m=\Lambda$.

As shown in \cite{Qi}, the cMERA state Eq.~\eqref{eq:APPevol} is a Gaussian state annihilated by the $a^{\Lambda}_s(k)$ of the form Eq.~\eqref{eq:app:aLams}, where
\begin{equation}
\label{eq:dalphak}
(\partial_s-k\partial_k)\alpha_s(k)=-2\alpha_s(k)g(k).
\end{equation}
Now we can substitute Eq.~\eqref{eq:app:aLams} into Eq.~\eqref{eq:dalphak} and see that it holds for arbitrary $s\in [0,\infty)$. Since $\alpha_s(k)$ uniquely determines a Guassian state, we have shown that $\ket{\Psi^{\cM}(s)}$ is in the ground state of the massive free boson Hamiltonian with a UV cutoff $\Lambda$ and mass $m=\Lambda e^{-s}$.

\subsection{Appendix II: computation of correlations functions}

Next we will compute correlation functions, involving the bosonic fields $\psi(x),\psi^{\dagger}(x)$, of a Gaussian state annihilated by 
\begin{equation}
a^{\cM}(k) \equiv \sqrt{\frac{\alpha(k)}{2}} \phi(k) + \frac{i}{\sqrt{2\alpha(k)}}\pi(k)
\end{equation}
with some function $\alpha(k)$. 
 First, we express the $\psi(k)$ in terms of these annihilation operators:
\begin{equation}
\psi(k)=\sqrt{\frac{\Lambda}{2}}\phi(k)+\frac{i}{\sqrt{2\Lambda}}\pi(k),
\end{equation}
where
\begin{eqnarray}
\phi(k)=\frac{1}{\sqrt{2\alpha(k)}}(a^{\Lambda}(k)+a^{\Lambda\dagger}(-k)) \\
\pi(k)=-i\frac{\sqrt{2\alpha(k)}}{2}(a^{\Lambda}(k)-a^{\Lambda\dagger}(-k)).
\end{eqnarray}
The correlation functions can then be computed from the canonical commutation relations $[a^{\Lambda}(k),a^{\Lambda\dagger}(k')]=\delta (k-k')$,
\begin{eqnarray}
\langle \psi(k)\psi(k')\rangle &=& \frac{1}{4} \left(\frac{\Lambda}{\alpha(k)}-\frac{\alpha(k)}{\Lambda}\right) \delta(k+k')  \\
&\equiv& F(k) \delta(k+k'), \nonumber \\
\label{eq:APPcorr2}
\langle \psi^{\dagger}(k)\psi(k')\rangle &=& \frac{1}{4} \left(\frac{\Lambda}{\alpha(k)}+\frac{\alpha(k)}{\Lambda}-2\right) \delta(k-k') ~~~ \\
&\equiv& n(k) \delta(k-k'). \nonumber
\end{eqnarray}
Transforming them into real space, we obtain,
\begin{equation}
\langle \psi(x)\psi(y) \rangle = \langle \psi^{\dagger}(x)\psi^{\dagger}(y) \rangle = \int \frac{dk}{2\pi} F(k) e^{ik(x-y)},
\end{equation}
\begin{equation}
\label{nn}
\langle \psi^{\dagger}(x)\psi(y)\rangle =\int \frac{dk}{2\pi} n(k)  e^{ik(x-y)}.
\end{equation}
In particular, the particle density $\rho_0\equiv\langle \psi^{\dagger}(x)\psi(x)\rangle$ can be computed analytically with the fixed point $\alpha(k)$ in Eq.~\eqref{eq:APPalphak},
\begin{equation}
\frac{\rho_0}{\Lambda}=\frac{1}{8\pi}\int  dk\, \left(\sqrt{\frac{1+k^2}{k^2}}+\sqrt{\frac{k^2}{1+k^2}}-2\right).
\end{equation}
The above expression has an IR divergence at $k=0$. Introducing a small mass $m\ll\Lambda$, i.e.,
\begin{equation}
\alpha(k) = \Lambda \sqrt{\frac{k^2+m^2}{k^2+\Lambda^2}}
\end{equation}
we find that
\begin{equation}
\frac{\rho_0}{\Lambda} = \frac{1}{8\pi} \log \frac{\Lambda^2}{m^2}-\frac{1.22741}{8\pi}+O\left(\frac{m^2}{\Lambda^2}\right).
\end{equation}
During the evolution Eq.~\eqref{eq:APPevol}, the cMERA state is the ground state of the massive Hamiltonian with mass $m=\Lambda e^{-s}$. We therefore expect that the particle density $\rho_0$ increases linearly with $s$ for $s\gg 1$. This is a direct consequence of the IR divergence, which is a feature of the free boson CFT in 1+1 dimensions. Note that, however, the ground state energy density of the massless Hamiltonian
\begin{eqnarray}
e_0 &\equiv& \langle \partial_x\psi^{\dagger}\partial_x\psi\rangle+\frac{\Lambda^2}{2}\rho_0-\frac{\Lambda^2}{4}\langle\psi^2+\psi^{\dagger 2}\rangle \\
&=& \int \frac{dk}{2\pi} \, \left(k^2 n(k)+ \frac{\Lambda^2}{2} (n(k)-F(k))\right).
\end{eqnarray}
 is \textit{finite} because the IR divergences in $n(k)$ and $F(k)$ cancel each other. Its value only depends on the UV cutoff,
 \begin{equation}
 e_0=-\frac{\Lambda^2}{6\pi}.
 \end{equation}
The fact that $e_0$ is finite is important in the context of numerical optimizations of cMERA through energy minimization \cite{Martin}. 

\subsection{Appendix III: cMERA as the ground state of a local Hamiltonian: generic case}
Consider a Gaussian cMERA state annihilated by 
\begin{equation}
a^{\cM}(k) \equiv \sqrt{\frac{\alpha(k)}{2}} \phi(k) + \frac{i}{\sqrt{2\alpha(k)}}\pi(k)
\end{equation}
with some function $\alpha(k)$. Then it is the ground state of all Hamiltonians with the form
\begin{equation}
H = \int dk\, E(k) a^{\Lambda}(k)^{\dagger} a^{\Lambda}(k),
\end{equation}
where $E(k)\geq 0$ can be any dispersion relation. In terms of original fields, the Hamiltonian is
\begin{equation}
H=\frac{1}{2}\int dk \, \left(\frac{E(k)}{\alpha(k)} \pi(k)\pi(-k)+E(k)\alpha(k)\phi(k)\phi(-k)\right). ~~
\end{equation}
This Hamiltonian is local (that is, it involves only a finite number of derivatives of the field operators) and invariant under spatial parity only if
\begin{eqnarray}
\frac{E(k)}{\alpha(k)}&=&P_1(k^2) \\
E(k)\alpha(k)&=&P_2(k^2),
\end{eqnarray}
where $P_1$ and $P_2$ are (finite-degree) polynomials. Then we have
\begin{eqnarray}
E(k)&=&\sqrt{P_1(k^2)P_2(k^2)} \\
\label{eq:APPalphak0}
\alpha(k)&=&\sqrt{\frac{P_2(k^2)}{P_1(k^2)}}.
\end{eqnarray}

Following Ref. \cite{Qi}, we will also require that both the CFT dispersion relation and the CFT ground state be recovered at small $k$, that is
\begin{equation}
\label{eq:APPEkgeneral}
E(k)=|k|+o(k),
\end{equation}
and
\begin{equation}
\label{eq:APPalphakgeneral}
\alpha(k)=|k|+o(k),
\end{equation}
and that the cMERA state approaches the product state in the UV, that is
\begin{equation}
\label{eq:APPUVcutoff}
\lim_{k\rightarrow\infty} \alpha(k)=\Lambda.
\end{equation}
Expanding the polynomials $P_1$ and $P_2$ as
\begin{eqnarray}
P_1(k^2) =\sum_{l=0}^{l_m} a_l (k^2)^l,~~~~ P_2(k^2) =\sum_{l=0}^{l'_m} b_l (k^2)^l,~~~
\end{eqnarray}
Eqs.~\eqref{eq:APPEkgeneral},~\eqref{eq:APPalphakgeneral} imply $a_0=1,b_0=0,b_1=1$. Eq.~\eqref{eq:APPUVcutoff} forces $P_1$ and $P_2$ to have the same degree $2l'_m=2l_m$ and also that $b_{l_m}=\Lambda^2 a_{l_m}$. The most generic \textit{local} quadratic Hamiltonian that has a cMERA ground state is therefore 
\begin{equation}
H^{\Lambda}=\frac{1}{2}\int dx \, \left(\sum_{l=0}^{l_m} a_l (\partial^l_x \pi(x))^2+ \sum_{l=1}^{l_m} b_l (\partial^l_x \phi(x))^2\right)
\end{equation}
subject to the above constraints. The magic cMERA in the main text corresponds to the simplest solution (polynomials of smallest degree), namely $l_m=1$ with $P_1=1+\frac{k^2}{\Lambda^2}$ and $P_2=k^2$. Note that the degrees of $P_1$ and $P_2$ determine the order of derivatives appearing in the Hamiltonian. In the main text, we have second order derivatives in $H$, which are $(\partial_x \phi(x))^2$ and $(\partial_x \pi(x))^2$, in accordance with the degree of $P_1$ and $P_2$. Choosing larger $l_m$ corresponds to regulating the CFT Hamiltonian with higher derivative terms. 

The asymptotic behavior of $\alpha(k)$ at large $k$ determines UV properties of the cMERA state. For the set of cMERA states with $\alpha(k)$ in Eq.~\eqref{eq:APPalphak0}, that is, the set of cMERA states that can be the ground state of a local Hamiltonian, it is always true that 
\begin{equation}
\label{eq:APPalphaconv}
\frac{\alpha(k)}{\Lambda}=1+O\left(\left(\frac{\Lambda^2}{k^2}\right)^n\right)
\end{equation}
with some positive integer $n\geq 1$. To determine $n$, we first find the smallest $l_1$ such that $b_l=\Lambda^2 a_l$ for all $l_1\leq l\leq l_m$, then $n=l_m-l_1+1$ is the number of such coefficients. Since $b_{l_m}=\Lambda^2 a_{l_m}$, it is clear that $1\leq n\leq l_m$. Now we can show that the cMERA in previous works \cite{cMERA1,Qi} cannot be the ground state of a local Hamiltonian. Indeed, the previous cMERA proposals involve a function $\alpha(k)$ that converges faster than any polynomial at large $k$, contradicting Eq.~\eqref{eq:APPalphaconv}.

Eq.~\eqref{eq:APPalphaconv} has various implications on the correlation functions. First, Eq.~\eqref{eq:APPcorr2} implies that
\begin{equation}
n(k)=O\left(\left(\frac{\Lambda^2}{k^2}\right)^{2n}\right).
\end{equation}
For $n=1$, $n(k)\sim 1/k^4$ which is compatible with a generic bosonic cMPS. The minimial choice $n=l_m=1$ gives the magic cMERA state in the main text. More generally, if $b_{l_m}=\Lambda^2 a_{l_m}$ but $b_{l_m-1}\neq \Lambda^2 a_{l_m-1}$, then $n=1$ and the ground state is compatible with the cMPS in the UV. If $n>1$, the state is compatible with a subclass of cMPS that satisfies certain regularity conditions, which imposes constraints on cMPS variational parameters.

Now consider the implication on the real space correlation function 
\begin{equation}
n(x)\equiv \langle \psi^{\dagger}(\vec{x})\psi(0)\rangle.
\end{equation}
it has continous derivatives at $x=0$ up to $4n-2$ order. For example, the expectation value of the non-relativistic kinetic term $\langle\partial_x\psi^{\dagger}\partial_x\psi\rangle=-\partial^2_x n(0)$ is always finite. However, higher order derivatives diverge in the $n=1$ case. We have therefore seen that, by asking the cMERA state to be the ground state of a local Hamiltonian, we automatically have correlation functions with finite orders of smoothness. This is to be in contrast with previous cMERA proposals \cite{cMERA1,Qi}, where correlation functions are infinite-order differentiable.

The entangler $K$ that generates this class of cMERA as the fixed point wavefunctionals also differs from previous works. The fixed point $\alpha(k)$ is related to $g(k)$ in Eq.~\eqref{eq:APPK} by
\begin{equation}
g(k)=\frac{k\partial_k \alpha(k)}{2\alpha(k)}.
\end{equation}
Substituting Eq.~\eqref{eq:APPalphak0} into the equation above, we obtain
\begin{equation}
g(k)=\frac{k^2}{2} \frac{P_1(k^2)P'_2(k^2)-P'_1(k^2)P_2(k^2)}{P_1(k^2)P_2(k^2)}.
\end{equation}
Note that $g(k)$ decays no slower than $1/k^2$ at large $k$ because $P_1(k^2)P'_2(k^2)-P'_1(k^2)P_2(k^2)$ has a degree that is at most $2l_m-4$. We see that $g(k)$ decays polynomially. This implies that its Fourier transform $g(x)$ at $x=0$ is not smooth. For example, the magic cMERA corresponds to $g(x)\propto e^{-\Lambda |x|}$, which does not have first-order derivative at $x=0$. This is in constrast with the Gaussian entangler $g(x)\propto e^{-\sigma(\Lambda x)^2/4}$ which is smooth at $x=0$.

At $k=0$, $P_1(0)=a_0=1, P'_2(0)=b_1=1$ and $\lim_{k\rightarrow 0} P_2(k^2)/k^2=b_1=1$, together give that $g(k)$ is smooth at $k=0$. To see this, let us rewrite
\begin{equation}
g(k)=\frac{1}{2} \left(\frac{P'_2(k^2)}{P_2(k^2)/k^2}-\frac{k^2P'_1(k^2)}{P_1(k^2)}\right).
\end{equation}
Both $P_2(k^2)/k^2$ and $P_1(k^2)$ are polynomials which are nonvanishing at $k=0$. This ensures that $g(k)$ is infinite-order differentiable at $k=0$. The fact that $g(k)$ is smooth at $k=0$ implies that $g(x)$ decays at least exponentially at large $x$, which keeps the entangler $K$ quasi-local. We can also work out
\begin{equation}
g(k=0)=\frac{1}{2},
\end{equation}
which ensures that the scaling dimensions (eigenvalues of $L+K$) come out correctly \cite{Qi}.

In conclusion, we have exhaustively determined the class of Gaussian bosonic cMERA states that can be the ground state of a local quadratic Hamiltonian. They (i) are characterized by two polynomials, (ii) have correlation functions compatible with a cMPS or a subclass of cMPS in the UV, and (iii) are generated by a quasi-local entangler with $g(x)$ decaying at least exponentially at large $x$ but not smooth at $x=0$.

\subsection{Appendix IV: Conformal group and scaling operators}

\subsubsection{1.Relation to conformal group}

The scale invariant magic cMERA $\ket{\Psi^{\Lambda}}$ in the main text is the exact ground state of any Hamiltonian of the form
\begin{equation}
H[E(k)] = \int dk~~E(k) a^{\Lambda}(k)^{\dagger} a^{\Lambda}(k)
\end{equation}
where the magic cMERA annihilation operators $a^{\Lambda}(k)$ are fixed, namely
\begin{eqnarray}
a^{\Lambda}(k) &\equiv& \sqrt{\frac{\alpha(k)}{2}}\phi(k) + \frac{i}{\sqrt{2\alpha(k)}}\pi(k), \\
\alpha(k) &\equiv&  \sqrt{\frac{k^2\Lambda^2}{k^2+\Lambda^2}},
\end{eqnarray} 
but where for the quasi-particle energies $E(k)$ we can choose any positive function. 

Two specific choices of $E(k)$ stand up. One makes $H$ strictly local, the other one makes $H$ part of a quasi-local representation of the conformal algebra.

In this work we wanted the Hamiltonian $H^{\Lambda}$ to be local. This requires the  choice 
\begin{equation}
E^{\Lambda}(k) \equiv \sqrt{\frac{k^2}{\Lambda^2}(k^2 + \Lambda^2)}.
\end{equation}
In Ref. \cite{Qi} we studied instead the dispersion relation $E^{\CFT}(k)=|k|$, in which case the Hamiltonian $H_{q.l.}^{\Lambda} \equiv H[E^{\CFT}(k)]$ is quasi-local, but by construction has the same spectrum as the local, relativistic CFT Hamiltonian $H^{\CFT}$ in \eqref{eq:HCFT}. (Notice that in Ref. \cite{Qi}, the quasi-local Hamiltonian $H_{q.l.}^{\Lambda}$ was denoted $H^{\Lambda}$). What makes $H_{q.l.}^{\Lambda}$ interesting is that it is part of a quasi-local realization of the conformal algebra, as described in Ref. \cite{Qi}. In particular, $D^\Lambda \equiv L+K$ is a quasi-local realization of the dilation operator, and we have that $D^{\Lambda}$ and $H_{q.l.}^{\Lambda}$ obey the commutation relation
\begin{equation}
-i\left[D^{\Lambda},H_{q.l.}^{\Lambda} \right] = H_{q.l.}^{\Lambda},
\end{equation}
which are the same as the commutation relation of CFT dilation operator $D^{\CFT}$ and CFT Hamiltonian operator $H^{\CFT}$, namely $\left[D^{\CFT}, H^{\CFT}\right]$. That is, $H^{\Lambda}_{q.l.}$ is scale invariant (under the scale transformation generated by $D^{\Lambda} = L + K$). 

Instead, by requiring locality, which is of importance from a computational perspective \cite{Martin}, in this work we used a Hamiltonian $H^{\Lambda}$ that is not scale invariant, that is $[D^{\Lambda}, H^{\Lambda}] \not = 0$. We note, however, that since $H^{\Lambda}$ and $H^{\Lambda}_{q.l.}$ have the same eigenvectors (indeed, by construction $\left[ H^{\Lambda}, H^{\Lambda}_{q.l.}\right] =0$) and their dispersion relations $E^{\Lambda}(k)$ of $E^{\CFT}(k)$ are very similar at low energies $k \ll \Lambda$, the violation of scale invariance is small at low energies.

\subsubsection{2.Derivation of scaling operators}

Following Ref. \cite{Qi}, the quasi-local scaling operators $\phi^{\Lambda}(x)$ and $\pi^{\Lambda}(x)$ are related to the sharp fields $\phi(x)$ and $\pi(x)$ by
\begin{eqnarray}
\phi^{\Lambda}(x) &=& \int dy\, \mu_{\phi}(x-y)\phi(y)\\
\label{eq:APPmupi}
\pi^{\Lambda}(x) &=& \int dy\, \mu_{\pi}(x-y)\pi(y),
\end{eqnarray}
where the Fourier transforms of the smearing functions are
\begin{eqnarray}
\mu_{\phi}(k)&\equiv& \sqrt{\frac{\alpha(k)}{|k|}}=\left(1+\frac{k^2}{\Lambda^2}\right)^{-1/2} \\
\mu_{\pi}(k)&\equiv& \sqrt{\frac{|k|}{\alpha(k)}}=\left(1+\frac{k^2}{\Lambda^2}\right)^{1/2}.
\end{eqnarray}
They have distributional Fourier transforms \cite{Qi}
\begin{eqnarray}
\mu_{\phi}(x)&=&\frac{2^{3/4}\Lambda K_{1/4}(|\Lambda x|)}{\Gamma(1/4)|\Lambda x|^{1/4}}\\
\mu_{\pi}(x)&=&\frac{2^{5/4}\Lambda K_{3/4}(|\Lambda x|)}{\Gamma(-1/4)|\Lambda x|^{3/4}}.
\end{eqnarray}
Note that Eq.~\eqref{eq:APPmupi} should be understood as the Hadamard finite-part integral
\begin{equation}
\pi^{\Lambda}(0)=\lim_{\epsilon\rightarrow 0} \left(2 \epsilon^{-1/2}\pi(0)+\int_{\mathcal{R}-(\epsilon,\epsilon)} dx\, \mu_{\pi}(x)\pi(x) \right).
\end{equation}
Other scaling operators include spatial derivatives $\partial^m_x \phi^{\Lambda}(x)$ with scaling dimensions $m$ and $\partial^m_x \pi^{\Lambda}(x)$ with scaling dimensions $m+1$. They can also be expressed as a distribution acting on the sharp fields $\phi(x),\pi(x)$, with profiles 
\begin{eqnarray}
\mu_{\partial^m_x\phi}(x) &=& \partial^m_x \mu_{\phi}(x)  \\
\mu_{\partial^m_x\pi}(x) &=& \partial^m_x \mu_{\pi}(x).
\end{eqnarray}
Some of the profile functions are plotted in the main text.

\subsection{Appendix V: Continuous matrix product operator}

\subsubsection{1. Matrix product operator (MPO)}
Consider a MPO made of matrices $A_m$ given by
\begin{equation}
A_m \equiv \left( \begin{array}{ccc}
\mathbb{1} & E_m & 0 \\
0 & \lambda \mathbb{1} & F_m \\
0 & 0 & \mathbb{1}
\end{array} \right),
\end{equation}
where $E_m$ and $F_m$ are two operators and $\mathbb{1}$ is the identity operator, all acting on the vector space of the lattice site $m$. 

\begin{widetext}
The product of two contiguous MPO matrices $A_m$ and $A_{m+1}$ is
\begin{eqnarray}
A_mA_{m+1} = \left( \begin{array}{ccc}
\mathbb{1} & E_m & 0 \\
0 & \lambda \mathbb{1} & F_m \\
0 & 0 & \mathbb{1}
\end{array} \right) \left( \begin{array}{ccc}
\mathbb{1} & E_{m+1} & 0 \\
0 & \lambda \mathbb{1} & F_{m+1} \\
0 & 0 & \mathbb{1}
\end{array} \right)
= \left( \begin{array}{ccc}
\mathbb{1} & \lambda E_m + E_{m+1} & E_mF_{m+1} \\
0 & \lambda^2 \mathbb{1} & F_m + \lambda F_{m+1} \\
0 & 0 & \mathbb{1}
\end{array} \right).
\end{eqnarray}

Similarly, the product $A_{m}A_{m+1}A_{m+2}$ reads
\begin{eqnarray}
A_{m}A_{m+1}A_{m+2} &=& \left( \begin{array}{ccc}
\mathbb{1} & \lambda E_m + E_{m+1} & E_mF_{m+1} \\
0 & \lambda^2 \mathbb{1} & F_m + \lambda F_{m+1} \\
0 & 0 & \mathbb{1}
\end{array} \right)
\left( \begin{array}{ccc}
\mathbb{1} & E_{m+2} & 0 \\
0 & \lambda \mathbb{1} & F_{m+2} \\
0 & 0 & \mathbb{1}
\end{array} \right)\\
&=&\left( \begin{array}{ccc}
\mathbb{1} & ~~\lambda^2 E_m + \lambda E_{m+1} + \lambda E_{m+2}~~ & 
~~E_{m} F_{m+1} + E_{m+1}F_{m+2} + \lambda E_m F_{m+2}~~\\
0 & \lambda^3 \mathbb{1} & F_m + \lambda F_{m+1} +\lambda^2 F_{m+2} \\
0 & 0 & \mathbb{1}
\end{array} \right),
\end{eqnarray}
and by iteration we find that the product $A_1 A_2\cdots A_N$ of $N$ such matrices reads
\begin{equation}
A_1 A_2\cdots A_N =\left( \begin{array}{ccc}
\mathbb{1} & ~~~\sum_{m=1}^N \lambda^{N-m} E_{m} ~~~& \sum_{m=1}^N \sum_{n=m+1}^{N} \lambda^{n-m-1} E_m F_n\\
0 & \lambda^{N} \mathbb{1} & \sum_{m=1}^N \lambda^{m-1} F_m \\
0 & 0 & \mathbb{1}
\end{array} \right).
\end{equation}
With the choice $E_m = F_m = \sqrt{\beta \lambda \epsilon}~ b_{m}$  and $\lambda = e^{-\epsilon\Lambda}$, the product becomes 
\begin{equation}
A_1 A_2\cdots A_N = \left( \begin{array}{ccc}
\mathbb{1} & ~~~ \sqrt{\beta\epsilon} ~\sum_{m=1}^N  e^{-\Lambda \epsilon(N-m+\frac{1}{2})} ~ b_{m} ~~~& \beta \epsilon~ \sum_{m=1}^N  \sum_{n=m+1}^{N}   e^{-\Lambda \epsilon(n-m)} ~b_{m} b_{n}~~\\
0 & e^{-\Lambda \epsilon N} \mathbb{1} & \sqrt{\beta\epsilon} ~\sum_{m=1}^N \lambda^{-\Lambda \epsilon(m-\frac{1}{2})}  ~ b_{m} \\
0 & 0 & \mathbb{1}
\end{array} \right).
\end{equation}
We are interested in the matrix element $(1,3)$ of this product, namely
\begin{equation}
\bra{1} A_1 A_2 \cdots A_N \ket{3} = \frac{-i\Lambda \epsilon}{4} \sum_{m<n} e^{-\Lambda \epsilon (n-m)} b_mb_n,
\end{equation} 
which accounts for one half of the discrete version $K^{\LAT}$ in the main text (the other half, quadratic in creation operators $b_m^{\dagger} b_n^{\dagger}$, is obtained similarly). 

\subsubsection{2. Continuous matrix product operator (cMPO)}

Next we introduce operators $\psi(x_m) \equiv b_m / \sqrt{\epsilon}$, where $x_m \equiv \epsilon m$, and expand the above matrix $A_m$ in powers of $\epsilon$,
\begin{equation}
A_m = \left( \begin{array}{ccc}
\mathbb{1} & E_m & 0 \\
0 & \lambda \mathbb{1} & F_m \\
0 & 0 & \mathbb{1}
\end{array} \right) = \left( \begin{array}{ccc}
\mathbb{1} &~~ \epsilon~\sqrt{\beta}e^{-\Lambda\epsilon/2}  \psi(x_m)& 0 \\
0 & e^{-\Lambda \epsilon} \mathbb{1} & ~~\epsilon~\sqrt{\beta}e^{-\Lambda\epsilon/2} \psi(x_m) \\
0 & 0 & \mathbb{1}
\end{array} \right) = \mathbb{1} + \epsilon \mathcal{A}_m + O(\epsilon^2 ),
\end{equation}
where the cMPO matrix $\mathcal{A}(x_m) = \mathcal{A}_{m}$ reads
\begin{equation}
\mathcal{A}(x_m) = \left( \begin{array}{ccc}
0 &~~ \sqrt{\beta} \psi(x_m) & 0 \\
0 & -\Lambda \mathbb{1}  & ~~\sqrt{\beta}\psi(x_m) \\
0 & 0 & 0
\end{array} \right).
\end{equation}
We can now expressed the matrix product $A_1 A_2 \cdots A_N$ in the double limit $\epsilon \rightarrow 0$ and $N\rightarrow \infty$, with finite $L = N\epsilon$, as a path ordered exponential,
\begin{eqnarray}
\mathcal{P}\exp \left(\int_0^L dx~\mathcal{A}(x) \right) \equiv \lim_{\small{\begin{array}{c} \epsilon\rightarrow 0\\N \rightarrow \infty \end{array}}} \left(1+\epsilon \mathcal{A}(x_1)\right) \left(1+\epsilon \mathcal{A}(x_2)\right)\cdots \left(1+\epsilon \mathcal{A}(x_N)\right),
\end{eqnarray}
whose matrix element $(1,3)$ reads
\begin{eqnarray}
\bra{1}\mathcal{P}\exp \left(\int_0^L dx~\mathcal{A}(x) \right)\ket{3} &=&  \frac{-i\Lambda}{4} \lim_{\small{\begin{array}{c}  \epsilon\rightarrow 0\\N \rightarrow \infty \end{array}}}\sum_{m=1}^N \epsilon \sum_{n=m+1}^N \epsilon~ e^{-\Lambda \epsilon (n-m)} \psi(x_m) \psi(x_n) \\
&=& \frac{-i\Lambda}{4} \int_0^L \!\!dx  \int_x^L \!\!dy ~ e^{- \Lambda|x-y|} \psi(x) \psi(y),
\end{eqnarray}
and thus accounts for half of the entangler $K$ in the main text.

\end{widetext}

We conclude that the entangler $K$ of the proposed magic cMERA can indeed be expressed in an extremely compact way using a cMPO. In Ref. \cite{Martin} this observation, which also implies a compact cMPO representation for $e^{isK}$ for small $s$, will be exploited as part of an efficient computational framework for cMERA, namely in order to numerically implement a scale evolution generated by $L+K$.


\begin{thebibliography}{99}
\bibitem{sigma} The smearing function $(\Lambda/4) \sqrt{\sigma/\pi}~e^{-\sigma(\Lambda x)^2/4}$ was proposed in Ref. \cite{cMERA1}, then corrected and further explored in Ref. \cite{Qi}. The constant $\sigma \approx 1.78107$ is the
exponential of Euler’s constant $\gamma \approx 0.57722$ and is needed for the resulting cMERA to be a good approximation to the free boson CFT ground state at large distances \cite{Qi}.
\bibitem{phiscaling} Strictly speaking, in 1+1 dimensions the field $\phi(x)$ is not a scaling operator of the free boson CFT, but $\partial \phi(x) \equiv \partial_x\phi(x) + i \pi(x)$ is. The same applies to the smeared fields $\phi^{\Lambda}(x)$ and $\partial \phi^{\Lambda}(x) \equiv \partial_x\phi^{\Lambda} (x) + i \pi^{\Lambda}(x)$.
\end{thebibliography}
\end{document}